\PassOptionsToPackage{activate={true,nocompatibility},final,tracking=true,kerning=true,spacing=true,factor=1100,stretch=10,shrink=10}{microtype}
\documentclass[sigconf,natbib=false,screen]{acmart}

\pdfoutput=1

\usepackage[utf8]{inputenc}
\usepackage[T1]{fontenc}

\usepackage{csquotes}
\usepackage[english]{babel}
\useshorthands*{"}
\defineshorthand{"-}{\babelhyphen{hard}}

\PassOptionsToPackage{sortcites,maxbibnames=99}{biblatex}
\usepackage{biblatex}
\bibliography{bibliography}

\usepackage{xcolor}
\usepackage{graphicx}
\usepackage{subcaption}
\usepackage{booktabs}
\usepackage{tabularx}
\usepackage{multirow}
\usepackage{makecell}
\usepackage{flushend}
\usepackage{tikz}
\usetikzlibrary{matrix}
\usepackage[htt]{hyphenat}

\usepackage{pgfplots}
\pgfplotsset{compat = 1.8}
\usepackage{pgfplotstable}
\pgfplotsset{compat=1.8,
    /pgfplots/ybar legend/.style={
    /pgfplots/legend image code/.code={%
       \draw[##1,/tikz/.cd,yshift=-0.25em]
        (0cm,0cm) rectangle (3pt,0.8em);},
   },
}
\usepackage{enumitem}
\setlist{nosep}

\usepackage{algorithm}
\usepackage[noend]{algpseudocode}
\algrenewcommand\algorithmicfunction{\textbf{func}}
\algrenewcommand\algorithmicprocedure{\textbf{program}}
\colorlet{AlgCommentGray}{black!80!white}

\usepackage[colorinlistoftodos,prependcaption]{todonotes}

\usepackage[binary-units]{siunitx}
\DeclareSIUnit{\nothing}{\relax}
\DeclareSIUnit{\x}{\times}
\DeclareSIUnit{\ct}{\text{\cent}}

\usepackage{xcolor}
\definecolor{ETHa}{RGB}{31,64,122}      %
\definecolor{ETHb}{RGB}{72,90,44}       %
\definecolor{ETHc}{RGB}{18,105,176}     %
\definecolor{ETHd}{RGB}{114,121,28}     %
\definecolor{ETHe}{RGB}{145,5,106}      %
\definecolor{ETHf}{RGB}{111,111,100}    %
\definecolor{ETHg}{RGB}{168,50,45}      %
\definecolor{ETHh}{RGB}{0,122,150}      %
\definecolor{ETHi}{RGB}{149,96,19}      %

\PassOptionsToPackage{pdfpagelabels=false}{hyperref}
\usepackage{hyperref}
\hypersetup{%
    colorlinks=true,%
    urlcolor=ETHg,%
    linkcolor=ETHc,%
    citecolor=ETHb,%
    breaklinks=true,%
    pageanchor=true,%
}

\newcommand{\jitq}{JITQ}

\makeatletter
\DeclareRobustCommand{\varname}[1]{\begingroup\newmcodes@\mathit{#1}\endgroup}
\makeatother

\title{The Collection Virtual Machine: \\
       An Abstraction for Multi-Frontend Multi-Backend Data Analysis}
\renewcommand{\shorttitle}{The Collection Virtual Machine}
\author{\vspace{-2.25ex}\hspace*{-2em}\mbox{%
    Ingo Müller$^1$\hspace{-.75ex},\hspace{.35ex}
    Renato Marroquín$^2$\hspace{-.75ex},\hspace{.35ex}
    Dimitrios Koutsoukos$^1$\hspace{-.75ex},\hspace{.35ex}
    Mike Wawrzoniak$^1$\hspace{-.75ex},\hspace{.35ex}
    Sabir Akhadov$^3$\hspace{-.75ex},\hspace{.35ex}
    Gustavo Alonso$^1$%
  }}
\renewcommand{\shortauthors}{
  I. Müller, R. Marroquín, D. Koutsoukos,
  M. Wawrzoniak, S. Akhadov, G. Alonso}
\affiliation{\vspace{-1.75ex}\small
  $^1$\texttt{\{ingo.mueller,dkoutsou,michal.wawrzoniak,alonso\}@inf.ethz.ch} \hfill
  $^2$\texttt{renato.marroquin@oracle.com} \hfill
  $^3$\texttt{sabir.akhadov@databricks.com}
}
\affiliation{\institution{\hspace{1.2em}
  Systems Group, Department of Computer Science, ETH Zurich \hfill
  Oracle Labs\hspace{2.5em}\hfill
  Databricks \hspace*{3.75em}
  \vspace{1ex}
}}

\acmConference[DaMoN'20]{Data Management On New Hardware}{June 15, 2020}{Portland, OR, USA}
\acmYear{2020}
\copyrightyear{2020}
\setcopyright{rightsretained}
\makeatletter
\def\@copyrightpermission{
  This paper is published under the Creative Commons
  Attribution-NonCommercial-ShareAlike license~4.0 International
  (CC~BY-NC-SA~4.0) license.
  To view a copy of this license,
  visit \url{http://creativecommons.org/licenses/by-nc-sa/4.0}.
}
\makeatother

\begin{document}

\begin{abstract}
Getting the best performance
from the ever-increasing number of hardware platforms
has been a recurring challenge for data processing systems.
In recent years, the advent of data science
with its increasingly numerous and complex types of analytics
has made this challenge even more difficult.
In practice, system designers are overwhelmed by the number of combinations
and typically implement only one analysis/platform combination,
leading to repeated implementation effort---%
and a plethora of semi-compatible tools for data scientists.

In this paper, we propose the ``Collection Virtual Machine'' (or CVM)---%
an extensible compiler framework designed
to keep the specialization process of data analytics systems tractable.
It can capture at the same time the essence of a large span
of low-level, hardware-specific implementation techniques
as well as high-level operations of different types of analyses.
At its core lies a language for defining nested, collection-oriented
intermediate representations (IRs).
Frontends produce programs in their IR flavors defined in that language,
which get optimized through a series of rewritings
(possibly changing the IR flavor multiple times)
until the program is finally expressed
in an IR of platform-specific operators.
While reducing the overall implementation effort,
this also improves the interoperability of both analyses and hardware platforms.
We have used CVM successfully to build specialized backends
for platforms as diverse as multi-core CPUs, RDMA clusters,
and serverless computing infrastructure in the cloud
and expect similar results
for many more frontends and hardware platforms in the near future.
\end{abstract}

\maketitle

\hypersetup{%
  pdfauthor={Ingo Müller and Renato Marroquín and Dimitrios Koutsoukos and Mike Wawrzoniak and Sabir Akhadov and Gustavo Alonso},
  pdftitle={The Collection Virtual Machine: An Abstraction for Multi-Frontend Multi-Backend Data Analysis},
}
\newcommand\blfootnote[1]{%
  \begingroup
  \renewcommand\thefootnote{}\footnote{#1}%
  \addtocounter{footnote}{-1}%
  \endgroup
}
\blfootnote{$^2$Most contributions of this author took place while affiliated with ETH Zurich.}
\blfootnote{$^3$The contributions of this author took place while affiliated with ETH Zurich.}

\vspace{-5ex}
\section{Introduction}
\label{sec:intro}

A major goal of systems design has always been
to translate increased hardware performance
into higher application performance.
This consists more and more
of exploiting specialized hardware across the entire stack---%
be it parallelization on the level of SIMD, multi-cores, NUMA, and machines,
or support for accelerators such as GPUs or FPGAs
or specialized IO devices
such as NVMe-based storage or RDMA-capable networking.
With the advent of data science
and its more diverse and more complex types of analytics,
the challenge for system designers has been extended to yet another dimension,
the support of multiple platforms at the same time.

While there is ample research
on how to exploit each hardware platform in isolation,
practitioners are struggling to build systems
that support more than one or a few of them at the same time.
Also, most popular Python packages for numerous types of analytics
with the size of datasets they support
as an indication of what platforms they run on.
Using RDBMSs and SQL can support datasets of virtually any size,
but the packages for data mining, linear algebra,
graph analytics, and machine learning
are mainly built for running on a single machine,
thus supporting datasets of at most some tens of gigabytes.
Systems scaling to racks or clusters, such as Spark~\cite{Zaharia2010},
do support larger datasets;
however, if (mis)used in a single-machine setup,
they are typically one or several orders of magnitude less efficient.
Overall, tools tend to specialize
in a relatively narrow type of analysis/platform combination.
As a consequence, individual users are forced to switch tools constantly
as their datasets, focus of investigation, or hardware change.
At the same time, many basic system components
are reimplemented in each of the specialized systems
leading to both higher implementation effort
and less efficient implementations.

To overcome that situation, this project aims to provide system designers with
a unified framework across both hardware platforms and target domains.
We hypothesize that
all (or at least most) modern hardware platforms
and types of data analysis used by data scientists today
are similar enough
to be expressed in intermediate representations (IRs)
based on the common abstraction of
\emph{(nested) transformations of (nested) collections}.
As we explain in more detail below,
analytics in relational algebra, graph analysis,
linear algebra, and machine learning
work on relations of records, set of vertices and edges,
vectors and matrices of numbers, and bags of samples, respectively,
all of which are some sort of ``collection'' of ``tuples'' of ``atoms''.
Also implementations, including the most optimized forms,
can be described naturally in a nested way:
inner loops can be seen
as the transformation of individual atoms of one collection into another
and the orchestration code around them
as the nested compositions of these transformations.

Based on this hypothesis, we build the ``Collection Virtual Machine''
(CVM),
a compiler framework for multi-frontend multi"-backend data analysis.
Its core is a language for defining
collection-oriented intermediate representations (IRs)
that consists of arbitrary collection-based ``instructions''
that we call ``operators.''%
\footnote{We use the terms interchangeably in this document.}
Frontends languages then map to a program in an IR defined in this language,
typically using high-level operators
that may be in part specific to that frontend.
Similarly, the backend of a particular hardware platform
can define its instructions expressing
the low-level implementation techniques required to maximize performance.
Since programs at all levels of abstraction
are expressed in the same IR language,
rewritings between them can be implemented
in a common optimizer framework
to bring the input program
into an optimized, platform-specific form.

We have used CVM for the IRs of three different systems:
\emph{JITQ}~\cite{Akhadov2017},
specialized for multi-core CPUs,
\emph{Modularis}~\cite{Koutsoukos2020},
specialized for RDMA clusters,
and \emph{Lambada}~\cite{Muller2020},
specialized for serverless cloud functions.
While the three platforms are diverse
and require different, specific implementation techniques,
they not only share CVMs compiler infrastructure
but also the overwhelming majority of their IRs
and the rewritings through which they are compiled.
Furthermore, they share a generic Python frontend
allowing data scientists to change platforms seamlessly.
In experiments, we show that the multi-core and RDMA-based systems
are roughly on par with mature systems specialized for these platforms
while the cloud backend is up to an order of magnitude faster
and up to two orders of magnitude cheaper
than two commercial RDBMSs optimized for the same use case.

\vspace{-1.5ex}
\section{Related Work}
\label{sec:rel_work}

Our work draws heavy inspiration from relational database systems.
Consequently, all work on query optimization and execution techniques
are relevant because we are designing CVM
such that all of them could be implemented in its IR language.
This includes work from the 90s on relational algebra
on nested relations~\cite{Roth1988} and sequences~\cite{Ramakrsihnan1998},
as well as more recent efforts on array database systems~\cite{Brown2010}.
Similarly, there are large bodies of research
on domain-specific implementation techniques for
linear algebra~\cite{Goto2008}, machine learning algorithms~\cite{Witten2017},
and graph analysis algorithms~\cite{Doekemeijer2014}
and we design CVM such that it can also incorporate these techniques.

There has been numerous projects increasing the breadth of systems
using a mix of compiler and query optimization techniques.
For example, TensorFlow XLA~\cite{Leary2017} is a high-level compiler built
to support different computing platforms including accelerators.
To combine different types of analysis in one system,
systems like LaraDB~\cite{Hutchison2017} and AIDA~\cite{Dsilva2018}
integrate relational algebra with linear algebra in a single abstraction.
Raven~\cite{karanasos2019extending} uses an IR
that enables cross-optimization and integrated execution
of ML inference and relational queries.
To target even more domains,
Tupleware~\cite{Crotty2015} and Weld~\cite{Palkar2018} use
query optimization and just-in-time compilation
to run algorithms from different domains efficiently
but the first is restricted to optimizations possible on UDFs
and the IR of the latter is fundamentally limited to shared-memory systems.
SystemDS~\cite{boehm2019systemds} builds on SystemML's~\cite{boehm2016systemml}
compilation toolchain and can run a wide range of data science processes
on multiple backends, including local CPU/GPU and Spark.
While the above are built on some kind of IR,
they all have in common that their IR consists of a fixed set of instructions,
making it difficult to extend with further frontends and backends
and thus support a more narrow analysis/platform combination than CVM targets.
More recently, the MLIR~\cite{Lattner2020} compiler framework aims to provide
tools and abstractions for expressing, transforming, and composing of
a wide range of intermediate representations and compilation to
a broad range of hardware targets, including ML accelerators,
but with a focus on deep learning on GPUs and TPUs.

\vspace{-1.5ex}
\section{The Collection Virtual Machine}

\subsection{Architecture Overview}
\label{sec:architecture}

CVM purposefully defines a \emph{language} of intermediate representations
instead of a concrete IR with a fixed set of instructions.
It fixes \emph{how} instructions and collection types look like---%
not \emph{which} of them exist.
This allows both frontends and hardware-specific backends
to define the precise building blocks they need
and still evolve as hardware, applications,
and experience in IR design make progress.

\begin{figure}[t]
    \centering
    \includegraphics[width=.85\linewidth]{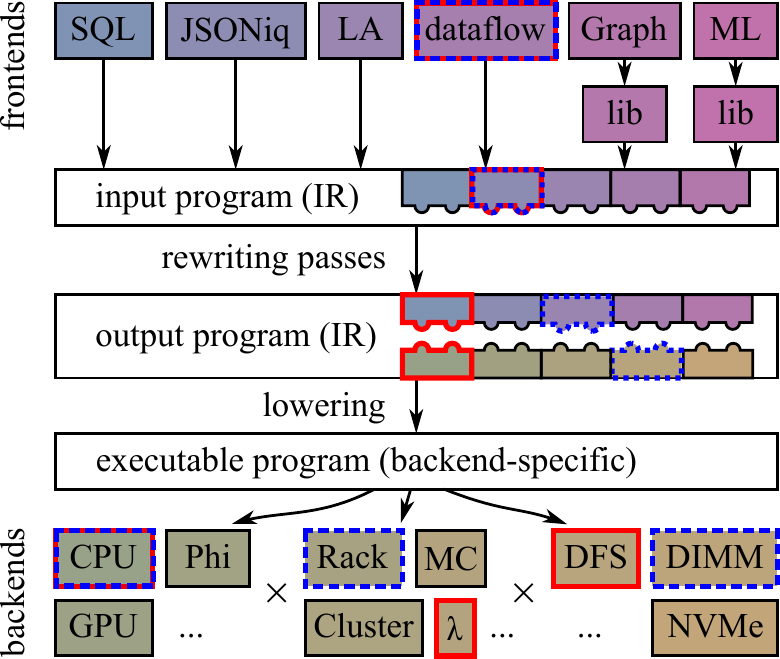}
    \caption{Architecture of the Collection Virtual Machine.
	Elements composing \emph{Lambada}~\cite{Muller2020}
	and \emph{Modularis}~\cite{Koutsoukos2020}
	are outlined in red and dashed blue lines respectively.}
    \label{fig:architecture}
\end{figure}

Figure~\ref{fig:architecture} illustrates an overview of the different components of the Collection Virtual Machine (CVM)
and the workflow of transforming a frontend program
into an executable form.
The figure shows several frontend languages and interfaces
that deal with collections,
but is not meant to be exhaustive.
Analyses in the frontends are expressed as or translated into
an intermediate representation (IR) defined in the CVM IR language.
This initial translation should be as thin as possible.
Frontends may define their own IR flavor
including high-level operators, collection types, or data types,
for example to perform domain-specific optimizations
in a frontend-specific rewriting pass.
The frontend and backends we implemented so far
are highlighted in the figure.

Once in a CVM IR, the program undergoes a succession of rewritings
that bring it into an optimized, executable form.
Which rewritings are applied and in which order
depends on the frontend and target backend(s) of the system.
During the rewriting, the program may change the IR flavor several times,
typically (but not necessarily)
going from more high-level IRs to more low-level ones
and intermediate programs may contain a mix of different IR flavors.
Since all programs use IRs defined in the same IR language,
mixing both IRs and rewritings is seamless
such that system builders can share their implementation efforts.
For example, the three systems we have implemented so far
share a common set of rewritings
that produce generic data-parallel programs in a common IR
and then rewrite some of the instructions
as instructions or sequences thereof
in their respective target-specific IR flavors.

Finally, the program is in a form
where its instructions correspond directly
to the executable building blocks of the target backend.
Like in compilers, we call the translation process of the final IR flavor
into that executable form \emph{lowering}.
For example, a traditional query compiler
would lower the IR of physical operators into an execution plan.
In JITQ, Modularis, and Lambada, we use a combination of two lowerings:
we lower pipelines representing the data paths
into native machine code using just-in-time compilation
and the surrounding orchestration logic
into a dataflow-based execution layer.

\subsection{IR Language}

All IRs in CVM are built on the mental model of an abstract virtual computer
that we call the ``Collection Virtual Machine''.
The virtual machine has an unlimited number of registers
that store ``collections''
and executes linear sequences of ``instructions'' called ``programs.''
Any transformation or execution of its IRs
must preserve the behavior \emph{as if it was executed on that machine}.

The IR language allows to define IR flavors
consisting of a set of instructions and collection types.
All collection types are generic with the following recursive structure:
\begin{equation}
  \varname{item} := \left\{\,
        \varname{atom} \,\middle|\,
        \varname{tuple}~\text{of}~\varname{items} \,\middle|\,
        \varname{collection}~\text{of}~\varname{items}
    \,\right\},
\end{equation}
where an \emph{atom} is an undividable value of a particular domain,
a \emph{tuple} is a mapping from a domain of names to items,
and a \emph{collection} is the generalization
of any (abstract or physical) data type
holding a finite, homogeneous multiset.
We denote tuples types by
$\langle\varname{fieldName0}: \varname{ItemType0},\,\ldots,\,
\varname{fieldNameK}: \varname{ItemTypeK}\rangle$
and collections types by
$\varname{CollectionType}\langle\varname{ItemType}\rangle$.

Instructions (or operators) defined by any IR flavor
have the following structure:
They read the collections
from zero or more previously assigned registers
and assign results to zero or more previously unassigned registers;
registers are hence immutable
and programs always in static single assignment (SSA) form.
Instructions may be parameterized with (constant) items and programs.
If an instruction takes a program as parameter,
we call it a \emph{higher-order} instruction.
Any instruction is thus of the following form%
\footnote{For brevity, we omit empty components.}:
\begin{algorithmic}
  \State $\varname{Out}_1$, \ldots, $\varname{Out}_m$
    $\gets$ \Call{Instruction}{$\varname{Para}_1$, \ldots, $\varname{Para}_k$}($\varname{In}_1$, \ldots, $\varname{In}_n$)
\end{algorithmic}
where $\varname{In}_i$ and $\varname{Out}_i$
are the input and output registers, respectively,
and $\varname{Para}_i$ the parameters (i.e., constant items and programs).

\subsection{Collection Types}

We now show how to define several collection types
in CVM's IR language to express
both abstract and physical data structures from various domains.
These examples are meant to show the expressiveness of the IR language
rather than a final set of types
and we expect to add more frontend and backend-specific types
as we implement other IRs in the future.
Table~\ref{tbl:collection-types}
shows the corresponding data types.

\begin{table}
  \centering
  \begin{tabular}{@{}l@{\hspace{1em}}l@{\hspace{1em}}l@{\hspace{-.5em}}}
    \toprule
    \textbf{Domain} & \textbf{Data structure} & \textbf{CVM data type} \\
    \midrule
    RA          & $R(A_1: D_1,\,\text{\ldots}\,A_k: D_k)$
                & $\varname{Set}\langle A_1: D_1,\,\text{\ldots}\,A_k: D_k\rangle$ \\
    Bag RA      & $R(A_1: D_1,\,\text{\ldots}\,A_k: D_k)$
                & $\varname{Bag}\langle A_1: D_1,\,\text{\ldots}\,A_k: D_k\rangle$ \\
    Seq. RA     & $R(A_1: D_1,\,\text{\ldots}\,A_k: D_k)$
                & $\varname{Seq}\langle A_1: D_1,\,\text{\ldots}\,A_k: D_k\rangle$ \\
    RA (NF$^\text{2}$)
                & $R_1(A: R_2(\text{\ldots}))$
                & $\varname{Bag}\langle A: \varname{Bag}\langle\text{\ldots}\rangle\rangle$ \\
                & $R(A_1: (A_2 : D))$
                & $\varname{Bag}\langle A_1: \langle A_2 : D\rangle\rangle$ \\
    LA          & $v \in \mathbb{R}$
                & $\varname{Seq}\langle\varname{Num}\rangle$ \\
                & $M \in \mathbb{R}^2$
                & $\varname{2DSeq}\langle\varname{Num}\rangle$ or \\
              & & $\varname{Seq}\langle\varname{Seq}\langle\varname{Num}\rangle\rangle$ \\
                & $M \in \mathbb{R}^k$
                & $\varname{kDSeq}\langle\varname{Num}\rangle$ or \\
              & & $\varname{Seq}\langle\text{\ldots}\,\varname{Seq}\langle\varname{Num}\rangle\text{\ldots}\rangle$ \\
    Graph       & $G = (V,E)$
                & $\varname{Set}\langle\varname{ID}\rangle$ and \\
              & & $\varname{Set}\langle\langle src: \varname{ID},\, dst: \varname{ID}\rangle\rangle$ \\
    \midrule
    row-store   & \texttt{struct\{ D1 field1; \ldots \}}
                & $\langle v1 : D_1,\,v2 : \text{\ldots}\rangle$ \\
                & \texttt{struct\{ D1 field1; \ldots \}*}
                & $\varname{Vec}\langle\langle v1 : D_1,\,v2 : \text{\ldots}\rangle\rangle$ \\
    col-store   & \texttt{struct\{ D1* col1; \ldots \}}
                & $\varname{Single}\langle\langle v1 : \varname{Vec}\langle\varname{D_1}\rangle,\,\text{\ldots}\rangle\rangle$ \\
    dense LA    & \texttt{float* A}
                & $\varname{Vec}\langle\varname{float}\rangle$ \\
                & \texttt{struct\{ int size[2];}
                & $\varname{Single}\langle\langle\varname{d1} : \varname{int},\,\varname{d2} : \varname{int},$ \\
                & \hspace{34pt}\texttt{float* A; \}}
                & \hspace{29pt}$A : \varname{Vec}\langle\varname{float}\rangle\rangle\rangle$ \\
    sparse LA   & \texttt{struct\{ int nnz;}
                & $\varname{Single}\langle\langle A : \varname{Vec}\langle\varname{float}\rangle,$ \\
                & \hspace{12pt}\texttt{float* A; int* I;}
                & \hspace{32.5pt}$\varname{I} : \varname{Vec}\langle\varname{int}\rangle,$ \\
                & \hspace{12pt}\texttt{int* O; \}}
                & \hspace{29pt}$\varname{O} : \varname{Vec}\langle\varname{int}\rangle\rangle\rangle$ \\
    SIMD        & \texttt{\_\_m256 v}
                & $\varname{Array8}\langle\varname{float}\rangle$ \\
    \bottomrule\addlinespace[.5ex]
    \multicolumn{3}{c}{\small
        $R$, $R_i$: relation; $A_i$, $A$: attribute/field name; $D_i$: atomic domain
        } \\
  \end{tabular}
  \smallskip
	\caption{Abstract (top) and physical (bottom) collection types.}
  \label{tbl:collection-types}
  \vspace{-3ex}
\end{table}

\textbf{Abstract collection types.}~~
First, collections can represent abstract data types,
as the ones shown in the upper half of the table.
To that aim, we define the generic collection types
\emph{Set}, \emph{Bag}, \emph{Seq} (for sequences),
and \emph{kDSeq} (for k-dimensional sequences).
We use them to compose high-level data types of various domains.
For example, relations from the original set-based relational algebra (RA)
are simply \emph{Set}s of tuples of atoms.
For this domain, the fact that items may be tuples is essential
as their field names and types represent the schema of the relation.
To express the more practical bag-based relations
(Bag RA, which is used in our current frontend),
sorted relations (from relational algebra on sequences, Seq. RA),
and relations in non-first normal-form (NF$^\text{2}$),
i.e., nested relations,
we simply use \emph{Bag} or \emph{Seq} instead of \emph{Set}
and allow non-atomic fields, respectively.
Similarly, we define vectors, matrices,
and higher"-dimensional tensors from linear algebra (LA),
as well as the vertices and edges of graphs.
In those domains, the items in the collections have no further structure,
but are just some type of number (\emph{Num})
or vertex identifier (\emph{ID}).

\textbf{Physical collection types.}~~
Second, collections can express physical data layouts as well.
As a basic building block, we define the generic type \emph{Vec} (for vector)
to represent an array of items in a single contiguous block of memory.
Furthermore, we express fixed-width records with ordered fields
(like \texttt{struct}s in C)
as tuples where the lexicographical order of the field names
defines the physical order in the layout.

This allows us again to compose many common physical data layouts,
such as those shown in the lower half of the table.
Both row-store and column-store layout of relations
are typically implemented as \emph{array of structs}
and \emph{struct of arrays},
which we can express with tuples and \emph{Vec}.
The three systems we have built so far use both relation types in their IRs.
Notice that we define the generic type \emph{Single}
as a singleton collection holding just one tuple
as a helper to store a group of collections in a single register.
Similarly, the data structures
used typically for linear algebra (both dense and sparse)
are composed of arrays and structs,
so we can express them with the same data types as shown in the table.
We only show the sparse matrix format CSR (for ``compressed sparse row''),
which consists of an integer for the number of non-zero elements
and three arrays (the non-zero elements, their column indices,
and the offsets of each row into the first two),
but the other common formats can be defined analogously.

Finally, as shown in the table, we define $\varname{ArrayN}$
as sequence with compile-time size $N$
to express vectors of machine words for SIMD-style processing.
The same collection type can also be used
to model a row of a narrow dense matrix
to enable compile-time optimizations for that special case.

The actual physical representation is decided by the lowering.
For example, our three systems
have an execution layer that stores tuples of fixed-width fields
in the memory layout of a C-arrays of C-structs
and thus require the final IR to contain
only \emph{Sequences} of anonymous tuples,
which are then lowered accordingly.
For that to work, we activate a sequence of rewriting passes
that bring the programs into the expected form.
We discuss these rewritings in more detail later in this section.

\textbf{Custom collection types.}~~
Finally, we can define new \mbox{collection} types
to support arbitrary physical formats and data \mbox{structures.
For} example, we have defined collection types for Apache Arrow and Parquet,
and other formats such as Protocol Buffers or Avro
could be supported with the same approach.
This allows frontends to support existing data formats
and backends to use specialized, highly"-tuned data structures
as data types in their respective IR flavor.

\vspace{-1.5ex}
\subsection{Instructions}
\vspace{-.5ex}

\begin{table}
  \centering
  \begin{tabular}{@{}l@{\hspace{1em}}l@{\hspace{1em}}l@{}}
    \toprule
    \textbf{Instruction} & \textbf{Input type(s)} & \textbf{Output type(s)} \\
    \midrule
    \textsc{Proj}($A_1,\,\text{\ldots}\,A_k$)($C$)
                    & $C : \varname{Coll}\langle A_1,\,\text{\ldots}\,A_k\,\text{\ldots}\rangle$
                    & $\varname{Bag}\langle A_1,\,\text{\ldots}\,A_k\rangle$ \\
                    & $C : \varname{Set}\langle A_1,\,\text{\ldots}\,A_k\,\text{\ldots}\rangle$
                    & $\varname{Set}\langle A_1,\,\text{\ldots}\,A_k\rangle$ \\
                    & $C : \varname{Seq}\langle A_1,\,\text{\ldots}\,A_k\,\text{\ldots}\rangle$
                    & $\varname{Seq}\langle A_1,\,\text{\ldots}\,A_k\rangle$ \\
    \textsc{ExProj}($\{A'_i, f_i\}_{l}$)($C$),
                    & $C : \varname{Coll}\langle A_1,\,\text{\ldots}\,A_k\rangle$
                    & $\varname{Bag}\langle \{A'_i : I_i\}_{l}\rangle$ \\
        \quad\small $f_i : \{A_j\}_{k} \rightarrow I_i$
                    & & \\
    \textsc{Map}($f : I_1 \rightarrow I_2$)($C$)
                    & $C : \varname{Coll}\langle I_1\rangle$
                    & $\varname{Bag}\langle I_2\rangle$ \\
                    & $C : \varname{Seq}\langle I_1\rangle$
                    & $\varname{Seq}\langle I_2\rangle$ \\
    \textsc{MMMult}($C_1$, $C_2$)
                    & $C_i : \varname{2DSeq}\langle Num\rangle$
                    & $\varname{2DSeq}\langle Num\rangle$ \\
    \midrule
    \textsc{Loop}($n$, $P$)($C_1$, \ldots\,$C_k$)
                    & $C_i : \varname{Coll}_i\langle I_i\rangle$
                    & $C_i : \varname{Coll}_i\langle I_i\rangle$ \\
        \quad\small $P : \{C_i\}_k \rightarrow \{C_i\}_k$
                    & & \\\addlinespace[.5ex]
    \textsc{While}($P$)($C_1$, \ldots, $C_k$)
                    & $C_i : \varname{Coll}_i\langle I_i\rangle$
                    & $C_i : \varname{Coll}_i\langle I_i\rangle$ \\
        \quad\small $P : \{C_i\}_k \rightarrow \mathbb{B}, \{C_i\}_k$\hspace*{-.2em}
                    & & \\\addlinespace[.5ex]
    \textsc{Cond}($P$)($C_1$, \ldots, $C_k$)
                    & $C_i : \varname{Coll}_i\langle I_i\rangle$
                    & $C'_j : \varname{Coll}'_j\langle I'_j\rangle$ \\
        \quad\small $P : \{C_i\}_k \rightarrow \mathbb{B}, \{C'_j\}^2_l$\hspace*{-.2em}
                    & & \\\addlinespace[.5ex]
    \textsc{Call}($P$)($C_1$, \ldots, $C_k$)
                    & $C_i : \varname{Coll}_i\langle I_i\rangle$
                    & $C'_j : \varname{Coll}'_j\langle I'_j\rangle$ \\
        \quad\small $P : \{C_i\}_k \rightarrow \{C'_j\}^2_l$
                    & & \\\addlinespace[.5ex]
    \textsc{ConcurExecute}($P$)($C$)
                    & $C : \varname{Coll}\langle I_1\rangle$
                    & $\varname{Bag}\langle I_2\rangle$ \\
        \quad\makecell[l]{\small $P : \varname{Single}\langle I_1 \rangle$ \\
            \hspace*{10pt}\small $\rightarrow \varname{Single}\langle I_2 \rangle$}
                    & \makecell[lt]{\vspace{-4ex}\\$C : \varname{Seq}\langle I_1\rangle$}
                    & \makecell[lt]{\vspace{-4ex}\\$\varname{Seq}\langle I_2\rangle$} \\[-.2ex]
    \midrule
    \textsc{ScanVec}(C)
                    & $C : \varname{Coll}\langle \varname{Vec}\langle I\rangle\rangle$
                    & $\varname{Seq}\langle I\rangle$ \\
    \textsc{MatVec}(C)
                    & $C : \varname{Coll}\langle I\rangle$
                    & $\varname{Single}\langle \varname{Vec}\langle I\rangle\rangle$\hspace*{-.5em} \\
    \textsc{SplitVec}($n$)(C)
                    & $C : \varname{Coll}\langle \varname{Vec}\langle I\rangle\rangle$
                    & $\varname{Bag}\langle \varname{Vec}\langle I\rangle\rangle$ \\
                    & $C : \varname{Seq}\langle \varname{Vec}\langle I\rangle\rangle$
                    & $\varname{Seq}\langle \varname{Vec}\langle I\rangle\rangle$ \\
    \textsc{BuildHTable}(C)
                    & $C : \varname{Coll}\langle T\rangle$,
                    & $\varname{Single}\langle \varname{HTab}\langle T\rangle\rangle$\hspace*{-.5em} \\
                    & $T : \langle key : I_1, val: I_2\rangle$
                    & \\
    \textsc{ProbeHTable}(C, H)
                    & $C : \varname{Coll}\langle T_1\rangle$,
                    & $\varname{Bag}\langle T_3\rangle$\hspace*{-.5em} \\
                    & $H : \varname{Single}\langle \varname{HTab}\langle T_2\rangle\rangle$,
                    & \\
                    & with $T_1.\varname{key} = T_2.\varname{key}$
                    & \\
    \bottomrule\addlinespace[.5ex]
    \multicolumn{3}{c}{\hspace{-1em}\makecell{
        \small$A_i$: attribute/field name; $C$, $C_i$: collection type; $I$, $I_i$: item type; $n\in\mathbb{N}$; \\
        \small$f$, $f_i$: function type; $P$: nested program; $\mathbb{B} = \{\top,\bot\}$; $T$, $T_i$: tuple type
        }\hspace*{-1em}} \\
  \end{tabular}
  \smallskip
  \caption{Domain-specific, control-flow-like, and low-level instructions.}
  \label{tbl:instructions}
  \vspace{-5ex}
\end{table}

As described above, instructions defined in the CVM IR language
transform collections into other collections.
Instructions may have restrictions on the item types of their input collections
and the types of their outputs may depend on their input types.
Broadly speaking, the level of abstraction of instructions
corresponds to the level of abstraction of the collections they work on.
Table~\ref{tbl:instructions} shows instructions
and their input and output types
of various levels of abstractions.

\textbf{High-Level Instructions.}~~
The upper part of the table
shows high-level, domain-specific instructions
that typically constitute the IRs used
for the initial translation of user-facing programs.
For example, an IR for a relational query processor
could define the usual relational operators on this level.
The table shows the definition of projection (\textsc{Proj}),
which is only defined on collections of tuples.
\footnote{However, the fields of the tuples may consist of arbitrary items.}
If the collection is a sequence (\emph{Seq}) or set (\emph{Set}),
then so is the output.
While the projection only restricts the field names of the tuples,
the extended projection (\textsc{ExProj}) also allows us to compute new fields.
We also define a \textsc{Map} instruction,
which we use in our generic dataflow frontend
and which, in contrast to the projections, can work on arbitrary item types.
We define instructions for other relational or generic dataflow operators
in much the same way as \textsc{Proj} and \textsc{Map};
however, we do not show them here due to space constraints.

As an example for an IR of a different application domain,
the table shows an instruction
for matrix-matrix multiplication (\textsc{MMMult}).
We can define instructions for other basic operations of linear algebra
including multiplications of tensors of different dimensions,
inversion, transposition, etc. analogously.
This allows to do high-level optimizations
based on mathematical and other domain-specific equivalences.

Notice how our definition of collections on different types of items
allows to express linear algebra and relational algebra in the same framework.
We can convert collections of one domain to the other
by packing (or unpacking) each item into (from) a tuple with a single field,
so our IR allows to combine programs of various frontends
and do optimizations across interface barriers.

\textbf{Control Flow.}~~
The middle part of the table
shows instructions we use to express control-flow-like behavior.
Notice that the CVM IR language
does not allow for traditional control flow such as jumps.
This is done by design
as jumps make it hard to understand the semantics of a program,
which makes many optimizations difficult or impossible to achieve.
However, we can use the capability of defining higher-order instructions
to achieve similar effects:
The table gives the example of a \textsc{Loop} instruction
that is parameterized with a nested program and a constant number $n$
and executes the program $n$ times.
It reads its input through input registers as any other instruction,
forwards them as initial input of the inner program,
and then uses the result registers of the previous run as new input.
The final result of the \textsc{Loop} instruction
corresponds to what the \textsc{Return} instruction
of the last run of the inner program returns.
\textsc{While} and \textsc{Cond} (for conditional expression)
can be defined in a similar way.%

\setlength\textfloatsep{3ex}

\begin{algorithm}[h]
  \begin{algorithmic}[1]
    \Statex\textcolor{AlgCommentGray}{\small$p$: predicate on ``l\_shipdate'', ``l\_discount'', and ``l\_quantity''}
    \Statex\textcolor{AlgCommentGray}{\small$T_{\varname{lineitem}}$: tuple corresponding to \emph{lineitem} schema}
    \Procedure{TpchQ6Seq}{$\varname{lineitem} : \varname{Coll}\langle T_{\varname{lineitem}}\rangle$}
    \State $\varname{filtered} \gets$ \Call{Select}{$p$}($\varname{lineitem}$)
    \State $\varname{projected} \gets$ \par
    \hskip\algorithmicindent \Call{ExProj}{$\text{``l\_eprice''}\cdot\text{``l\_disc''} \rightarrow \text{``x''}$}($\varname{filtered}$)
    \State $\varname{result} \gets$ \Call{Aggr}{$(\text{``x''}, \varname{sum}) \rightarrow \text{``revenue''}$}($\varname{projected}$)
    \State \Call{Return}{$\varname{result}$}
    \EndProcedure
  \end{algorithmic}
  \caption{Initial CVM program of TPC-H Query 6.}
  \label{alg:tpch6-initial}
\end{algorithm}

\vspace{-3ex}

\begin{algorithm}[h]
  \begin{algorithmic}[1]
    \Procedure{TpchQ6Par}{$\varname{lineitem} : \varname{Coll}\langle T_{\varname{lineitem}}\rangle$}
    \State $\varname{parts} \gets$ \Call{Split}{$p$}($\varname{lineitem}$)
    \State $\varname{part\_results} \gets$ \Call{ConcurExecute}{\textsc{TpchQ6Seq}}($\varname{parts}$)
    \State $\varname{unnested} \gets$ \Call{Scan}{$\varname{part\_results}$}
    \State $\varname{result} \gets$\par
    \hskip\algorithmicindent \Call{Aggr}{$(\text{``revenue''}, \varname{sum}) \rightarrow \text{``revenue''}$}($\varname{unnested}$)
    \State \Call{Return}{$\varname{result}$}
    \EndProcedure
  \end{algorithmic}
  \caption{Parallelized CVM program TPC-H Query 6.}
  \label{alg:tpch6-parallel}
\end{algorithm}

Parallel execution may also be counted as control flow.
The table shows the \textsc{ConcurrentExecute} instruction,
which we use to represent parallelism in our three systems.
It has similar semantics as the higher-order instruction \textsc{Map},
i.e., it executes a program on each input item to compute an output item,
but guarantees that these programs are executed concurrently
such that the different executions can exchange data among them.
Furthermore, each system has its own, platform-specific version
of \textsc{ConcurrentExecute},
which implements the concurrent execution of threads,
MPI workers, and serverless cloud functions, respectively.

\textbf{Low-Level Instructions.}~~
Finally, low-level instructions represent
specific building blocks of different backends.
On this level, we follow the philosophy to make these operators
as small as possible to make them more generic and hence reusable.
Our goal is to express cleverness
as a sophisticated combination of simple operators
instead of a simple combination of sophisticated operators.
We refer to our work on Modularis~\cite{Koutsoukos2020} for details.
For example, we have a scan operator, a materialize operator,
and potentially a split operator (for parallelization)
for each of the backend-level collection types mentioned above
(the table shows those of \emph{Vec}).
Similarly, we define a build and a probe operator
for each hash table type that we implement
(some of which are tuned for a very specific case).

Furthermore, many if not all low-level tuning techniques
developed by the database community in the last years
can be encapsulated as operators:
\begin{itemize}
  \item hardware-conscious algorithms and data structures,
  \item light-weight compression schemes,
  \item build and probe of for spatial indices or other domains, and
  \item predicated or vectorized scans, to name just a few.
\end{itemize}
Notice that all of them fall into our structure of CVM-based instructions.
Our IR language thus makes it possible
to use very specialized implementation techniques
and still represent them in a common abstraction.

\vspace{-1ex}
\subsection{Lowerings to Execution Layers}

The CVM compilation toolchain can be used to target any execution layer.
As mentioned before, a traditional relational query engine
could define its physical query plans as an IR
and lower programs in that IR
into an execution plan composed of its executable operators.
For example, we use MonetDB's execution layer
by translating programs through a series of rewritings
into an IR that replicates the MonetDB Assembly Language (MAL)
and then lowering them into actual MAL for execution.

For JITQ, Modularis, and Lambada,
we use a common execution layer for the data paths
and specialized components
for the execution of and communication among parallel workers.
The common part deals with the most fine-grained level,
where operators pass individual tuples between each other.
In a rewriting pass,
we extract tree-shaped parts of a program
and translate them into pipelines of Volcano-style iterators.
To eliminate the overhead of this operator interface
and to allow low-level optimizations across operator boundaries,
we just-in-time-compile each pipeline to native machine code.
The inputs and outputs of each pipeline
constitute necessary materialization points of the original program.

\subsection{Rewritings}

The rewriting mechanism of CVM is highly flexible and configurable,
such that every frontend/backend combination
can do the rewritings that are best suited for that combination.
For the different IR flavors to co-exist, at least during compilation,
rewritings must work
in presence of collection types and instructions of any IR.
Optimizations (or lowerings) that require a particular property
(such as tree-shaped data dependencies)
thus either have to rewrite the program to establish that property first
or work only on those parts of a program where the property holds.

Algorithms~\ref{alg:tpch6-initial} and~\ref{alg:tpch6-parallel}
illustrate how the rewriting for generic parallelization works
taking Query~6 of the TPC-H benchmark as an example.
The initial program (\textsc{Tpch6Seq}) consists of a selection,
a computation, and a scalar aggregation.
Our rewriting rule first replaces
the usage of the input relation (\emph{lineitem})
with a \textsc{Split}
followed by an empty \textsc{ConcurrentExecute} and a \textsc{Scan}.%
\footnote{This intermediate program is not shown.}
Notice that the sequence of these three operators is a logical no-op.
Then it applies rules that expand the \textsc{ConcurrentExecute}
in a way that preserves the semantics:
It moves \textsc{Select} and \textsc{ExProj} inside,
while it copies \textsc{Aggr} as pre-aggregation.
If an unknown instruction had been encountered,
then the rule would leave it as is.
The resulting parallelized program
is shown by Algorithm~\ref{alg:tpch6-parallel}.%
\footnote{The inner program of the \textsc{ConcurrentExecute}
          happens to be the same as the original program,
          which is why we refer to \textsc{Tpch6Seq}
          instead of spelling it out.}
As mentioned earlier, our three systems
each continue with a target-specific rewriting pass
that rewrite the program in Algorithm~\ref{alg:tpch6-parallel}
into an IR for thread-parallelism, RDMA clusters,
or cloud functions, respectively.

In the future, we plan to extend the rewritings considerably.
We think that all traditional query optimization techniques
from database systems can be done on CVM IRs,
including join reordering, index selection, etc.

\section{Experiments}
\label{sec:evaluation}

In this section, we show the experimental results of CVM on three
different hardware platforms: in-memory, distributed and serverless. Although
for some hardware configurations our platform is slower for analytical
workloads, the goal of the experimental study is not to focus on raw performance
but rather to show the flexibility of our frontends and backends through the use
of
platform-specific operators and rewrite rules. That comes with a reasonable
performance overhead compared to state-of-the-art data processing systems in
most cases, while there are some others where CVM is on par or even faster than
state-of-the-art. For all experiments described below, unless otherwise stated,
we run each query four times, use the first run as a warm-up and then report
the average of the other runs.

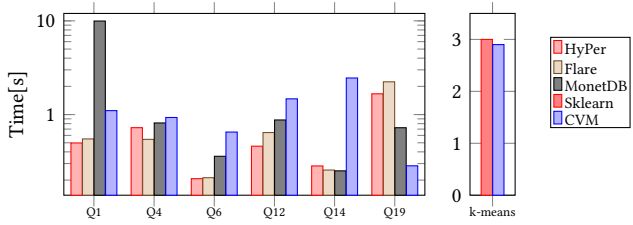
\begin{figure}
\centering
\begin{tikzpicture}
\begin{axis}[
    name=ax1,
    ybar,
    x=0.8cm,
    enlarge x limits={abs=0.4cm},
    ymax=12,
    ytick = {0, 0.1, 1, 10},
    legend style={at={(0.45,-0.15)},
    anchor=north,legend columns=-1, font=\tiny},
    ylabel={Time[s]},
    symbolic x coords={Q1, Q4, Q6, Q12, Q14, Q19},
    ylabel style={font=\small},
    xtick=data,
    bar width=0.15cm,
    ybar=0.1pt,
    inner ysep=0.01pt,
    nodes near coords align={vertical},
    x tick label style={font=\tiny,text width=1cm,align=center},
    y tick label style={font=\small},
    ymode=log,
    log origin=infty,
    log ticks with fixed point,
    yminorticks=true,
    height=4cm,
    width=7cm,
    ]
    \addplot[color=red, fill=red!30!white] coordinates {(Q1, 0.498) (Q4, 0.725) (Q6, 0.207) (Q12, 0.460) (Q14, 0.283) (Q19, 1.666)};
    \label{plots:hyper}
    \addplot[color=brown!60!black, fill=brown!30!white] coordinates {(Q1, 0.55) (Q4, 0.544) (Q6, 0.212) (Q12, 0.643) (Q14,
0.256) (Q19, 2.236)};
    \label{plots:flare}
\addplot[color=black, fill=gray] coordinates {(Q1, 9.959) (Q4, 0.814) (Q6, 0.359) (Q12, 0.876) (Q14,
0.251) (Q19, 0.724)};
    \label{plots:monetdb}
\addplot[color=blue, fill=blue!30!white] coordinates {(Q1, 1.102) (Q4, 0.931) (Q6, 0.651) (Q12, 1.472) (Q14,
2.457) (Q19, 0.284)};
    \label{plots:cvm}
\end{axis}
\begin{axis}[
    name=ax2,
    at={(ax1.south east)},
    xshift=0.6cm,
    ybar,
    x=0.2cm,
    enlarge x limits={abs=0.1cm},
    ymin=0,
    ymax=3.5,
    legend style={at={(0.45,-0.15)},
    anchor=north,legend columns=-1, font=\tiny},
    ylabel style={font=\small},
    xtick=data,
    symbolic x coords={k-means},
    bar width=0.15cm,
    ybar=0.1pt,
    inner ysep=0.01pt,
    nodes near coords align={vertical},
    x tick label style={font=\tiny,text width=1cm,align=center},
    y tick label style={font=\small},
    height=4cm,
    width=7cm,
    ]
    \addplot[color=red, fill=red!50!white] coordinates {(k-means, 3)};
    \label{plots:sklearn}
\addplot[color=blue, fill=blue!30!white] coordinates {(k-means, 2.9)};
\end{axis}
    \matrix[
        matrix of nodes,
        draw,
        font=\scriptsize,
        inner xsep=0.5pt,
        inner ysep=0.5pt,
        column 2/.style={nodes={anchor=base west}},
        align=left,
        ]at(7, 1.5)
      {
        \ref{plots:hyper}& HyPer\\
        \ref{plots:flare}& Flare\\
        \ref{plots:monetdb}& MonetDB\\
        \ref{plots:sklearn}& Sklearn\\
        \ref{plots:cvm}& CVM\\};

\end{tikzpicture}
\caption{TPC-H (SF 10) and k-means on a single machine.}
\label{fig:single-tpch}
\end{figure}

\textbf{In-memory.} \
For the in-memory experiments, we use two workloads: (1) TPC-H queries with
scale factor 10, and (2) the k-means clustering algorithm with a synthetic
dataset comprising of $2^{24}$ 5-dimensional points.
The numbers for HyPer and Flare are taken from~\cite{Essertel2018}.
For k-means, we choose the
most popular ML Python package, \texttt{scikit-learn} (``sklearn''),
as a competitor and we report the time of a single iteration.
Both experiments were run on an Intel Xeon E5-2630 v3 CPU 2.4 GHz.
We report the execution times for TPC-H queries (left) and the k-means
algorithm (right) in Figure~\ref{fig:single-tpch}.

We observe that the column-wise operations performed by MonetDB for Q1
have a negative impact in the running time whereas \jitq~lowers the same query
into a single pipeline which leads to producing the result in a single pass.
We also observe that when the input data are largely reduced due to very
selective filters,
such as in Q19, \jitq~outperforms the competitors.
We believe that implementing other missing optimizations like support
for narrow data types, a sophisticated optimizer and
index-based grouping will make our performance on par
for other queries as well.

For k-means, we achieve the performance of the hand-written C++ library used
under the hood in \texttt{scikit-learn}, mainly due to a plan analysis that
enables run-based aggregation.  The experiments show how that high-level
analysis and just-in-time-compilation achieves in-memory processing
speed that matches hand-written code performance.

\textbf{Distributed RDMA cluster.} \
For the distributed experiments, we use 8 machines each with two CPUs Intel
Xeon E5–2609 2.40 GHz and 128GB of RAM\@. The machines are connected through an
InfiniBand network with a Mellanox QDR HCA network card. We use TPC-H queries
with scale factor 500 and compare against two popular distributed systems,
MemSQL and Presto which were configured to use the entire cluster. For
Presto, we use HDFS nodes with default configurations to store TPC-H data.

\begin{figure}
\centering
\begin{tikzpicture}
\begin{axis}[
    ybar,
    x=1.4cm,
    enlarge x limits={abs=0.7cm},
    ymin=0,
    ymax=60,
    legend style={at={(1.2, 0.7)},
    anchor=north,legend columns=1, font=\scriptsize, inner xsep=1pt,
    inner ysep=1pt},
    ylabel={Time[s]},
    ylabel style={font=\small},
    ytick={0, 2, 4, 8, 50},
    extra y ticks = {1, 3, 5, 6, 7},
    extra y tick labels=\empty,
    legend cell align={left},
    symbolic x coords={Q4, Q12, Q14, Q19},
    xtick=data,
    legend entries={MemSQL, CVM w/o reading, Presto, CVM w reading},
    bar width=0.2cm,
    ybar=0.1pt,
    inner ysep=0.1pt,
    nodes near coords align={vertical},
    x tick label style={font=\scriptsize,text width=1cm,align=center},
    y tick label style={font=\small},
    every node near coord/.append style={font=\tiny},
    ymode=log,
    log ticks with fixed point,
    yminorticks=true,
    height=4cm,
    width=8cm,
    ]
\addplot[color=red, fill=red!30!white] coordinates {(Q4, 5.13) (Q12, 3.24) (Q14, 1.88) (Q19, 2.13)};
\addplot[color=blue, fill=blue!30!white] coordinates {(Q4, 5.3985892) (Q12, 3.4290368) (Q14, 2.7957672) (Q19, 2.847793)};
\addplot[color=brown!60!black, fill=brown!30!white] coordinates {(Q4, 48.82867) (Q12, 47.60967) (Q14, 34.43933) (Q19, 56.17333)};
\addplot[color=blue, fill=blue!60!white] coordinates {(Q4, 7.4359014) (Q12, 6.3527884) (Q14, 5.7808142) (Q19, 6.6066768)};
\end{axis}
\end{tikzpicture}
\vspace{-3ex}
\caption{TPC-H (SF 500) on an RDMA cluster.}
\label{fig:tpch-modularis}
\end{figure}
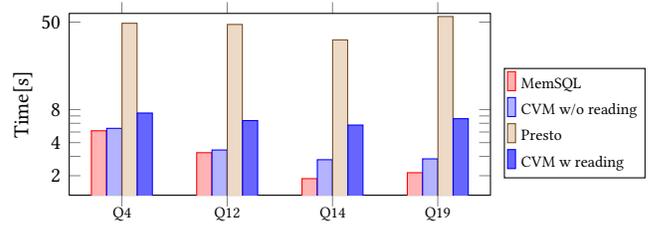

Figure~\ref{fig:tpch-modularis} shows the running times for executing the TPC-H
queries across the three systems.  To have a fair comparison with Presto, we
also include
the time that Modularis needs to read the input data.  We observe that for
queries 4 and 12 Modularis is on par with MemSQL and regarding queries 14 and
19, MemSQL is 33\% and 25\% faster, respectively.  Our system is 6-9x faster
than Presto, depending on the query.
Therefore, we can conclude that Modularis' performance is very close to a highly
optimized in-memory distributed database and orders of magnitude faster than a
popular big-data SQL query engine.

Additionally, in contrast to the above specialized systems, CVM supports this
platform,
only by implementing a few hardware-con-scious operators (i.e.,
\texttt{MPIExecutor}, \texttt{MPIExchange}, \texttt{MPIHistogram})
and by adding additional rewrite rules for
incorporating such operators. For instance, we wrote a specialized version of
\texttt{ConcurrentExecute} called \texttt{MPIExecutor} that uses OpenMPI to distribute
processes among the machines in the cluster.

\begin{figure}[H]
  \centering
  \begin{tikzpicture}
\begin{axis}[
    name=ax1,
    ybar,
    x=0.5cm,
    enlarge x limits={abs=0.25cm},
    ymin=0,
    ymax=26,
    legend style={at={(2.55, 0.7), font=\scriptsize},
    anchor=north,legend columns=1, align=left, inner xsep=0.5pt,
    inner ysep=0.1pt},
    legend cell align={left},
    ylabel={Time[s]},
    ylabel style={font=\scriptsize},
    ybar=0.1pt,
    inner ysep=0.1pt,
    ytick={0, 5, 10, 15, 20, 25},
    symbolic x coords={Q1, Q4, Q6, Q12, Q14, Q19},
    xtick=data,
    bar width=0.1cm,
    nodes near coords align={vertical},
    x tick label style={font=\tiny,text width=1cm,align=center},
    y tick label style={font=\tiny},
    every node near coord/.append style={font=\tiny},
    height=4cm,
    width=8cm,
    ]
    \addplot[color=red, fill=red!30!white] coordinates {(Q1, 25.15) (Q4, 23.74) (Q6, 12.9) (Q12, 17.55) (Q14,
    16.26) (Q19, 21.34)};
    \addplot[color=black, fill=gray] coordinates {(Q1, 6.01) (Q4, 13.69) (Q6, 0.77) (Q12, 5.54) (Q14,
    3.19) (Q19, 2.99)};
    \addplot[color=blue, fill=blue!30!white] coordinates {(Q1, 4.28) (Q4, 16.69) (Q6, 3.69) (Q12, 13.59) (Q14,
    10.26) (Q19, 10.39)};
    \legend{Athena, BigQuery, CVM}
\end{axis}
\begin{axis}[
    at={(ax1.south east)},
    xshift=0.9cm,
    name=ax2,
    ybar,
    x=0.5cm,
    enlarge x limits={abs=0.25cm},
    ymin=0,
    ymax=1.8,
    ybar=0.1pt,
    inner ysep=0.1pt,
    ylabel={Cost[$\$$]},
    ylabel near ticks,
    ylabel style={font=\scriptsize},
    ytick={0, 0.2, 0.5, 1, 1.5, 1.7},
    symbolic x coords={Q1, Q4, Q6, Q12, Q14, Q19},
    xtick=data,
    bar width=0.1cm,
    nodes near coords align={vertical},
    x tick label style={font=\tiny,text width=1cm,align=center},
    y tick label style={font=\scriptsize},
    every node near coord/.append style={font=\tiny},
    height=4cm,
    width=8cm,
    ]
    \addplot[color=red, fill=red!30!white] coordinates {(Q1, 0.27) (Q4, 0.19) (Q6, 0.24) (Q12, 0.23) (Q14,
    0.40) (Q19, 0.40)};
    \addplot[color=black, fill=gray] coordinates {(Q1, 1.68) (Q4, 0.90) (Q6, 0.96) (Q12, 1.31) (Q14,
    0.98) (Q19, 1.47)};
    \addplot[color=blue, fill=blue!30!white] coordinates {(Q1, 0.02) (Q4, 0.21) (Q6, 0.02) (Q12, 0.16) (Q14,
    0.13) (Q19, 0.13)};
\end{axis}

\end{tikzpicture}
  \caption{TPC-H (SF 1k) on serverless platforms.}
  \label{fig:tpch-all-queries}
\end{figure}
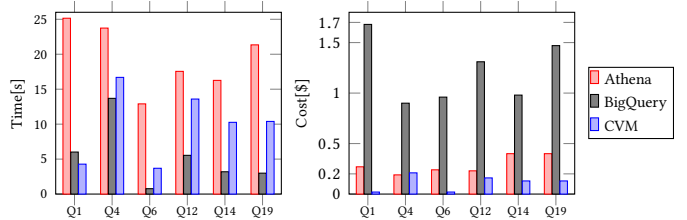

\textbf{Serverless functions.} \
Finally, we show how Lambada executes analytical workloads on a serverless
platform. Figure~\ref{fig:tpch-all-queries} shows the running time and monetary
cost of TPC-H queries on serverless cloud services. To showcase the
elasticity of serverless computing, we use TPC-H data at scale factor 1000 and
decide to use as many serverless workers needed to enable running queries
with \textit{interactive latencies}.
Compared to other serverless solutions, Google BigQuery and Amazon Athena,
Lambada is up to an
order of magnitude faster and up to two orders of magnitude cheaper. This shows
that the addition of new lowerings for this platform is orthogonal to
existing optimizations.

It is worth restating that when using CVM, the same frontend programs run
as in the previous experiments while only adding operators and incorporating
rewriting rules. For instance, Lambada lowers \texttt{ConcurrentExecute} into
\texttt{ParallelLambdaMap} an operator that
invokes AWS Lambda workers, and
it also transforms other operators into Amazon S3 specific operators.
Other optimizations such as selections and
projections can still be applied by putting them directly
into the operator that reads from Amazon S3. Adding such
functionality in the other serverless solutions would possibly imply major code
rewrites.

\section{Conclusions}
\label{sec:conclusions}

In this paper, we proposed the Collection Virtual Machine,
an abstraction for system designers
that keeps supporting the growing number of combinations
of domain-specific frontends and hardware backends tractable.
We have used CVM for the IRs of three different systems:
\emph{JITQ}~\cite{Akhadov2017},
\emph{Modularis}~\cite{Koutsoukos2020},
and \emph{Lambada}~\cite{Muller2020}.
While their target platforms are diverse,
we have shown how CVM allows the three systems
to share large parts of their IRs and rewritings in a common framework
and still get comparable performance
with systems designed from scratch for the respective hardware platforms.
In the near future, we plan to add other frontends
as well as more hardware platforms,
where we expect similar results.

\printbibliography

\end{document}